# Enhancing Light Extraction of Organic Light Emitting Diodes by Deep-Groove High-index Dielectric Nanomesh Using Large-area Nanoimprint

*Ji Qi, Wei Ding, Qi Zhang, Yuxuan Wang, Hao Chen, and Stephen Y. Chou\**

\*Prof. S. Y. Chou
Department of Electrical Engineering
Princeton University
Princeton, New Jersey 08544, USA
E-mail: chou@princeton.edu

J. Qi, W. Ding, Q. Zhang, Y. Wang, H. Chen
Department of Electrical Engineering,
Princeton University
Princeton, New Jersey 08544, USA

Keywords: light extraction, high-index, deep-grove, power efficiency, organic light emitting diode, nanostructures, nanoimprint, scattering efficiency

**Abstract**

To solve the conventional conflict between maintaining good charge transport property and achieving high light extraction efficiency when using micro/nanostructure patterned substrates to extract light from light emitting diodes (OLEDs), we developed a novel OLED structure, termed "High-index Deep-Groove Dielectric Nanomesh OLED" (HDNM-OLED), fabricated by large-area nanoimprint lithography (NIL). The key component is a nanostructure-patterned substrate embedded with a high-index deep-groove nanomesh and capped with a low-index planarization layer. The high-index and deep-groove nanomesh efficiently releases the tapped photons to achieve significantly enhanced light extraction. And the planarization layer helps to maintain the good charge transport property of the organic active layers deposited on top of it. For a green phosphorescent OLED in our demonstration, with the HDNM-OLED structure, compared to planar conventional ITO-OLED structure, the external quantum efficiency (EQE) was enhanced by 1.85-fold from 26% to 48% and power efficiency was enhanced by 1.86-fold from 42lm/W to 79lm/W. Besides green OELDs, the HDNM-OLED structure was also shown to be able to work for red and blue-emitting OELDs and achieved enhanced light extraction efficiency (1.58-fold for red light, 1.86-fold for blue light) without further structure modification, which demonstrated the light extraction enhancement by the HDNM-OLED is broadband.

**Introduction**

Organic light emitting diodes (OELDs) have attracted considerable attention as a promising technology for display and lighting applications due to their high efficiency, easy fabrication, low driving voltage and ability for making on flexible substrates. And many researches nowadays are being conducted in industrial and academic fields to address the key technical issues of potential OLED application. One of the most important driving forces is to enhance the light extraction. Because there is a large refractive index difference between air and the OLED materials (substrate and organic active layers), even though current phosphorescent OLED's internal quantum efficiency (IQE) has approached to 100%[1-2], the light extraction efficiency of the OLEDs with typical planar structures is only around 20% to 30%[3-6]. Around 80% of the generated photons in the OLEDs are trapped in the device and finally convert to the heat loss. To overcome this problem, different approaches in device design have been proposed such as using microlens array, high-index substrate, microcavity and micro/nano-structure structure patterned substrate. Among those methods, microlens array can only extract the light trapped in substrate mode; the microcavity and high-index substrate either require special device layer thickness design or new material development. Compared to the other methods, the micro/nano-structure patterned substrate is a more promising and universal way for extracting light from OLEDs because it utilizes scattering to extract the light which less depends on the specific device layer design and materials.

Different microstructure or nanostructure patterned substrate have been developed and studied by researchers, including micro/nanomesh and nanopillars. Although many efforts and progress have been made in this field, currently one of the key problem of using micro/nanostructure-patterned substrate for light extraction in OLEDs is that there is a trade-off between achieving high light extraction enhancement and maintaining good charge transport property[7-8]. Micro/nanostructure-patterned substrates rely on the scattering to extract the trapped photons, and therefore to achieve high light extraction enhancement, according to the scattering theory, the micro/nanostructures must have enough large size and height to efficiently scatter the light[20-22]. However, the thin organic light emitting layers which are typically around 100nm thick cannot sustain such large surface roughness. The OLEDs fabricated on such high-roughness substrates will have poor charge transport properties which will degrade its intrinsic photon generating efficiency (i.e., internal quantum efficiency)[23-25]. Previous studies from other researches have observed this effect. The researches A.O. Altun *et al* used nanopillars patterned substrates to enhance the light extraction efficiency of green-emitting OLEDs with 140nm-thick organic layers and they found that when the pillar height was larger than 50nm, the light extraction enhancement began to drop and even dropped below the light extraction efficiency of the control device fabricated on a planar substrate after the pillar height went above 100nm[7]. Similar effect was also observed in our previous work of DNM-OLED when we used dielectric-nanomesh patterned substrate to enhance light extraction of red-emitting OLEDs: the maximum light extraction was achieved at the groove depth of 40nm and further increasing the groove depth in turn degraded the efficiency due to the poor charge transport property of the organic layers under large corrugation. To solve this conflict, we developed a new OLED structure, named high-index deep-groove dielectric-nanomesh OLED (HDNM-OLED) that has a nanostructure patterned substrate embedded with a high-index deep-groove nanomesh and capped with a low-index planarization layer, which significantly enhances the light extraction without electrical degradation. Moreover, the HDNM-OLED substrates we developed were fabricated by nanoimprint lithography (NIL) rather than other complicated and time-consuming methods such as laser interference lithography, electron beam lithography (EBL) and focused ion beam lithography (FIB), and hence easily to applied in real large-area manufacture process with high throughput.

**HDNM-OLED structure, design and principle.** The HDNM-OLED was fabricated on a novel designed HDNM-substrate which is the key component providing the enhanced performance in HDNM-OLED. The HDNM-substrate comprises a subwavelength nanomesh pattern etched on a glass substrate, a high-index deep-grove (300nm) nanomesh on top of the glass substrate, a top low-index planarization layer to planarize the nanomesh (Fig. 1). On top of the HDNM-OLED substrate, 100nm-thick indium-tin-oxide (ITO) front transparent electrodes, 140nm-thick organic active layers and 0.5nm/140nm-thick lithium fluoride (LiF)/aluminum (Al) back electrodes are sequentially deposited to form the final HDNM-OLED device (Fig. 1). As we can see from the device structure schematic in Fig. 1, the electrods and organic layers are very flat even though there are structures with large corrugations in the substrate, which ensures the good charge transport property of the HDNM-OLED. We first applied the HDNM-OLED structure to a green-emitting OLED to study the light extraction enhancement performance of the HDNM-OLED. In an optimized green-emitting HDNM-OLED structure, the nanomesh pattern on the glass substrate has a 400nm-pitch square lattice with square-hole arrays, and the line-width of 130nm and the groove depth of 300nm. The high-index material embedded between the nanomesh patterned glass and the top 200nm-thick spin-on planarization layer is 400nm-thick tantalum pentoxide ($Ta_2O_5$) with a large refractive index of ~2.2. The organic light emitting layers has a four-layer structure consisting of 40nm-thick N,N'-Bis(naphthalen-1-yl)-N,N'-bis(phenyl)benzidine (NPB) as the hole injection layer, 20nm Tris(4-carbazoyl-9-ylphenyl)amine (TCTA) as the hole transport layer, 30nm thick host material 4,4'-Bis(N-carbazolyl)-1,1'-biphenyl (CBP) doped with 6 wt% high efficient green-emitting phosphorescent guest molecules Tris[2-phenylpyridinato-C2,N]iridium(III) [Ir(ppy)$_3$], and 50nm-thick 2,2',2"-(1,3,5-Benzinetriyl)-tris(1-phenyl-1-H-benzimidazole) (TPBI) as the electron transport layer.

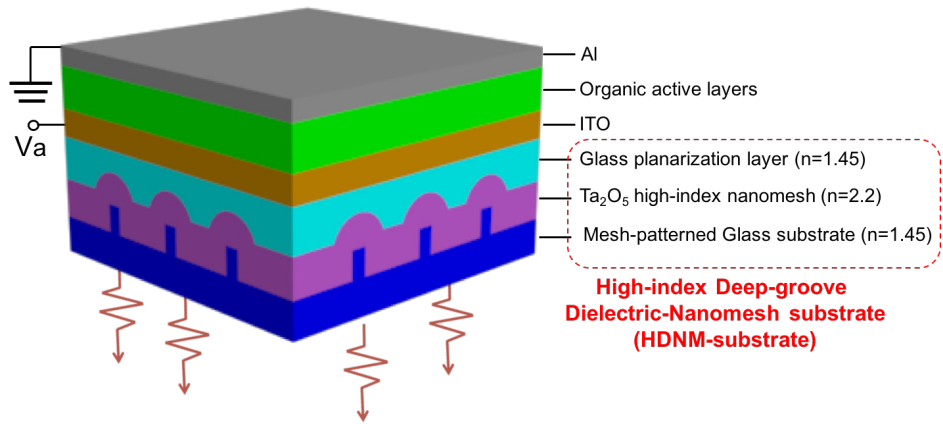

**Figure 1. High-index Deep-groove Dielectric-nanomesh Organic Light Emitting Diode (HDNM-OLED).** The layer structure schematic: a bottom high-index dep-groove dielectric-nanomesh substrate (HDNM-substrate) with sandwich structure consists of a nanomesh patterned glass substrate, a high-index nanomesh in the middle and a glass planarization layer on top; anode, organic active layers and cathode deposited on top of the substrate to form HDNM-OLED device.

The nanomesh structure in the HDNM-substrate is the key component to provide the light extraction enhancement property. It utilizes the Bragg scattering to extract the photons that are trapped in the waveguide and substrate mode. Therefore, to achieve a high light extraction enhancement, the nanomesh must be able to scatter the light efficiently. For the scatters with the size in the same order of magnitude as the wavelength, Mie scattering theory dominates the scattering behavior between scatterers and light[20-22]. And according to the Mie scattering theory, the effective scattering cross section (Q) is a function of the two parameters: ratio of refractive index between scatterers and medium ($\rho = \frac{n_s}{n_0}$); and the ratio of the size between scatterers and wavelength ($q = \frac{2\pi n_0 r_s}{\lambda}$):

$$Q\left(\frac{n_s}{n_0}, q\right) = \frac{2}{q^2} Re(\sum_{l=1}^{\infty}(-i)^{l+1}(l+1)(\mathcal{E}\left(\frac{n_s}{n_0}, q\right) + \mathcal{M}\left(\frac{n_s}{n_0}, q\right)))$$

where the above equation applies to scattering of a sphere in a medium.

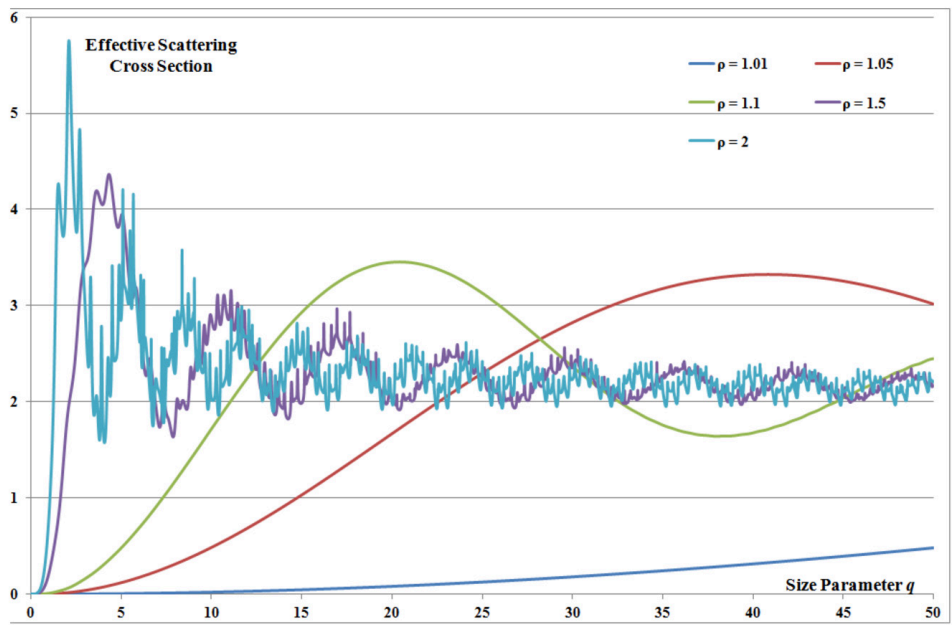

**Figure 2. The effective scattering cross section as a function of the scatterer size for different refractive index ratios upon Mie scattering theory.**

And the Fig. 2 is the plot of Q vs. q for several different ρ based on the above formula, from which we can see that generally, larger scatterer size and larger refractive index difference between scatterers and medium lead to higher scattering efficiency. Hence, a deep-groove (300nm) nanomesh pattern on the glass was designed to make the nanomesh features have the comparable size to the visible wavelength of light to ensure efficient light scattering. And a sandwich structure was further designed consisting of a high-refractive-index material ($Ta_2O_2$, n=2.2) layer deposited on top of the deep-groove nanomesh on glass (n=1.45) and capped with a low-index glass planarization layer (n=1.34) to make a large refractive index difference between the nanomesh scatterer and its medium to further increase its scattering efficiency. The top planarization layer not only provides a low-index medium for the high-index nanomesh scatterer but also provides a flat surface for the following OLED electrodes and organic layer deposition which ensure the charge transport property of HDNM-OLED as good as the conventional planar ITO OLED.

**Fabrication**

**HDNM-OLED fabrication**

The HDNM-substrate was fabricated by large-area nanoimprint lithography. The detailed fabrication process is illustrated in Fig. 2. It started with a 0.5mm-thick glass substrate with a NIL resist layer spin-coated on top of it. Then the nanomesh pattern was formed by hot embossing the 4-inch nanomesh NIL mold into the resist layer. After etching the residual resist layer, the nanomesh pattern is exposed on the glass substrate. Next, a chrome (Cr) nanomesh is formed on the glass substrate after Cr deposition and lift-off. By using the Cr nanomesh as an etching mask, the top 300nm-thick of glass substrate was etched off, forming the deep-groove nanomesh. After removing the Cr, 400nm-thick $Ta_2O_5$ was deposited to fully cover the nanomesh patterned glass substrate. In the end, 200nm-thick spin-on-glass (IC1-200, Futurrex, Inc) was used to planarize the top surface by annealing at 400°C after spin-coating.

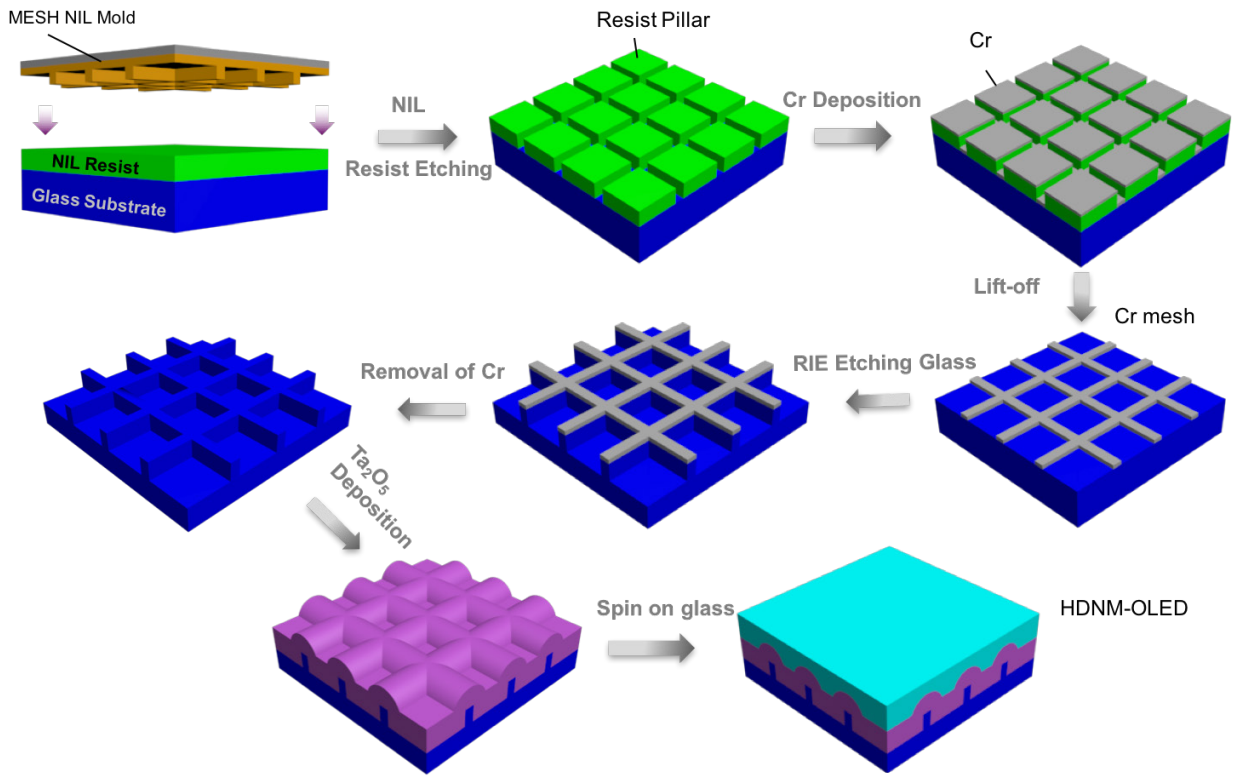

**Figure 2. Fabrication of high-index deep-groove dielectric nanomesh substrate (HDNM-substrate).** fabrication process: fabrication of nanomesh pattern in the resist layer by nanoimprint, Cr deposition and lift-off, RIE etching of glass substrate with Cr nanomesh as mask, removal of Cr, deposition of $Ta_2O_5$ and spin-on-glass planarization.

Scanning electron microscopy (SEM) images show that the deep-groove nanomesh patterned into the glass substrate has 400nm pitch, 120nm linewidth and groove depth of 300nm (Fig.3b). The top view SEM image confirms the excellent pattern uniformity of the nanomesh and the inset cross section image shows that the nanomesh has sharp edges and smooth sidewalls with good fidelity to the mold.

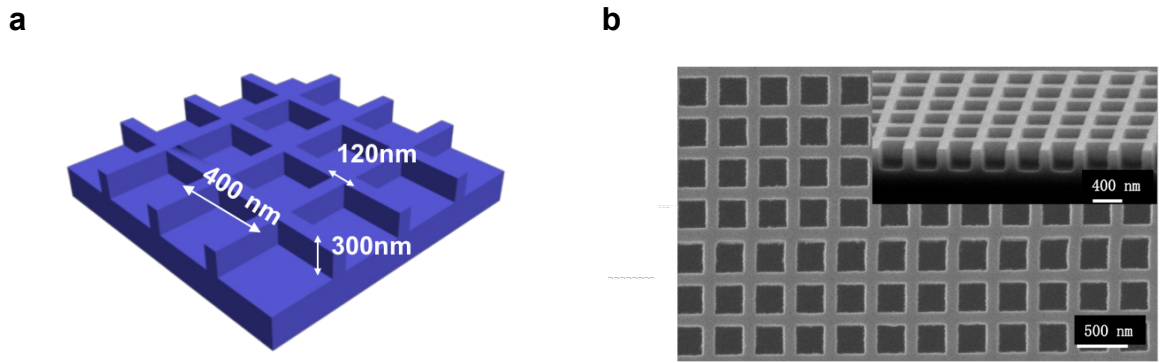

**Figure 3. Deep-groove nanomesh pattern on glass substrate.** (a) schematic of the deep-groove nanomesh pattern etched into glass substrate. (b) SEM image of the deep-groove nanomesh pattern in glass. Top view and cross section (inset).

The SEM image of the cross-section of HDNM-substrate is shown in Fig. 4b. It is clearly seen that the top surface is very flat and smooth as designed. One interesting thing needed to be pointed out is that, as shown in Fig. 4b, the $Ta_2O_5$ did not fill up the deep-groove nanomesh in glass substrate and left many vacuum pockets embedded between the glass substrate and $Ta_2O_5$ layer. This is explicable if considering the shadowing effect during the e-beam evaporation[26-27] (Fig. 5): the width of the $Ta_2O_5$ layer deposited on top of the higher parts of nanomesh grids gradually increased during the evaporation process and blocked the deposition of Ta$_2$O$_5$ to the lower plane of the nanomesh. Therefore, the Ta$_2$O$_5$ deposited in the holes of the nanomesh has triangle shapes as shown in the cross-section SEM image and there are vacuum pockets between the glass nanomesh and Ta$_2$O$_5$ layer. Although those vacuum pockets are not intentionally designed in the HDNM-OLED structure, they actually further help to improve the light extraction performance because they can act as additional scattering centers with high-refractive-index contrast.

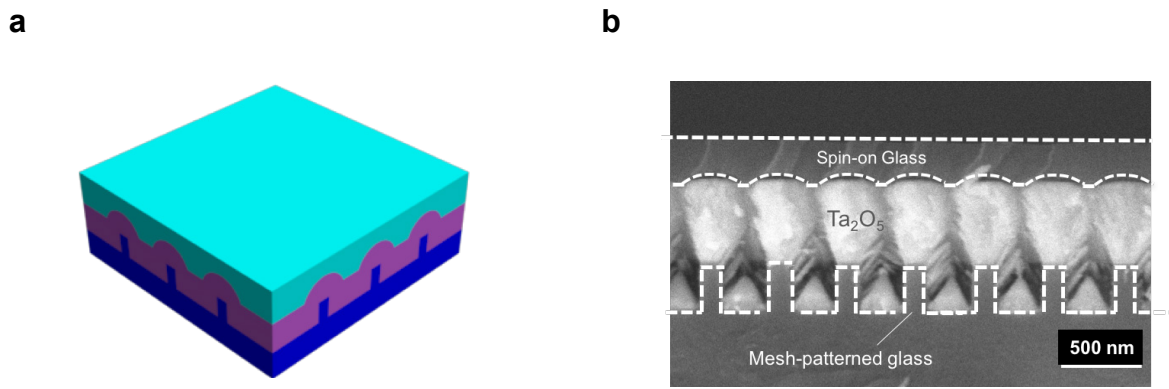

**Figure 4. High-index deep-groove nanomesh substrate (HDNM-substrate).** (a) schematic of HDNM-substrate. (b) SEM image of the cross-section of HDNM-substrate. White dashed lines indicate the interface between different layers.

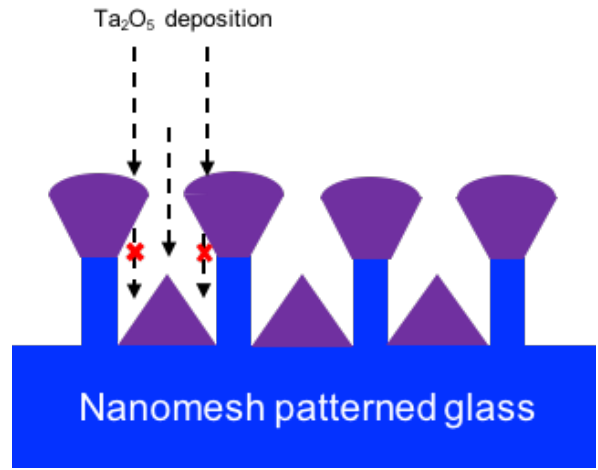

**Figure 5. Shadowing effect during Ta₂O₅ deposition.** The Ta$_2$O$_5$ deposited on the top of the higher grid bars gradually going wider during e-beam deposition and blocked the deposition to the bottom plane of the glass nanomesh.

And as shown in Fig. 6a, in the final step, the green-emitting HDNM-OLED was fabricated by sequentially depositing 140nm-thick ITO layer by e-beam evaporation, 40nm-thick NPB, 20nm-thick TCTA, 30nm-thick CBP doped with 6wt% Ir(ppy)$_3$, 50nm-thick TPBI and 0.5nm/100nm thick LiF/Al by thermal evaporation under high vacuum (<10$^{-7}$ Torr). The light emitting area of a HDNM-OLED is 3mm by 3mm, which is defined by a shadow mask during the evaporation of Al back electrode. The cross-sectional SME image (Fig. 6c) of the HDNM-OLED shows that the organic layers and electrode layers deposited on top of the HDMM-substrate are very uniform and flat.

**a**

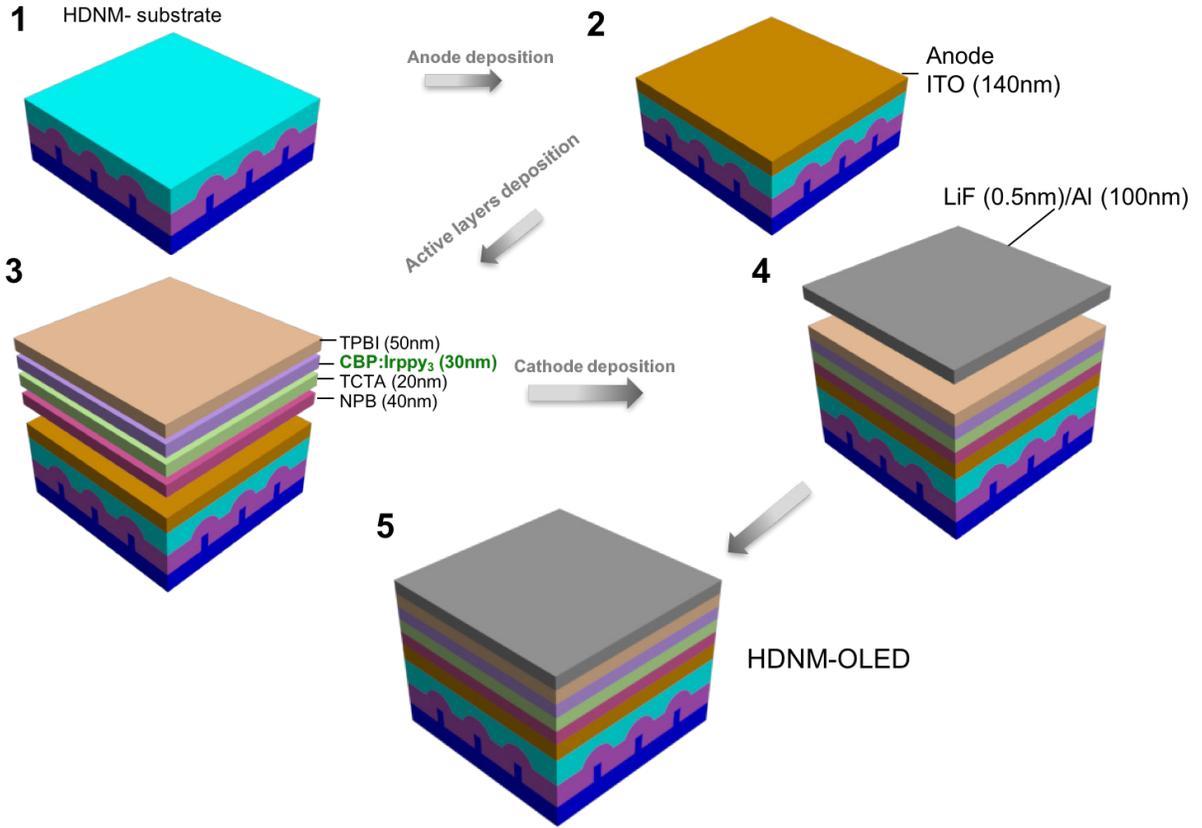

**b**

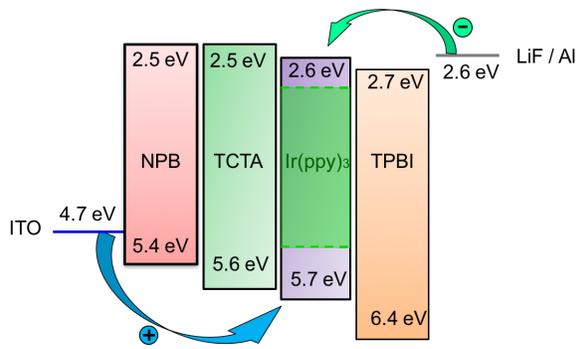

**c**

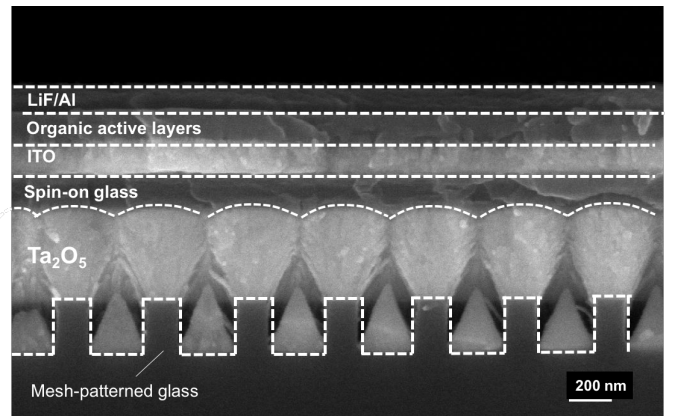

**Figure 6. Fabrication of the green-emitting HDNM-OLED.** (a) schematic of fabrication process: ITO anode, organic active layers and LiF/Al cathode were sequentially deposited on of top of the HDNM-substrate. (b) energy band diagram; (c) cross-sectional SEM image of the green emitting HDNM-OLED (white dashed lines indicate the interfaces between layers).

For comparison, reference OLEDs, "ITO-OLEDs" were also fabricated, which have the same organic active layer structure as the HDNM-OLEDs except replacing the HDNM-substrates with planar glass substrates.

**Results and discussion**

**Electroluminescence, EQE and light extraction enhancement.** The total front surface electroluminescence (EL) spectra of the green-emitting HDNM-OLED and ITO-OLED (Fig. 7a) were measured using an integrated sphere (Labsphere LMS-100) connected to a spectrometer (Horiba Jobin Yvon). During the measurement, the devices are attached on a stage holder in the integrated sphere and the side walls of the devices were fully covered by black tapes to ensure only the light emitted from OLEDs' front surfaces is collected by the spectrometer. The measured EL spectra show that in the entire emitting range (450nm to 620nm), the EL intensity of HDNM-OLED is much higher than that of ITO-OLED. And for the green-emitting material we use, the strongest emission intensity is achieved in the wavelength range of 520nm to 560nm. Compared with the ITO-OELD, at this emission peak, the HDNM-OLED shows a 1.75-fold EL enhancement factor (i.e. the ratio of the spectrum of HDNM-OLED to ITO-OLED) and by averaging the EL enhancement factor over the entire emission wavelength range, the HDNM-OLED shows a 1.71-fold average EL enhancement factor.

The EQE as a function of injection current density of the HDNM-OLEDs and ITO-OLEDs (Fig.

4b) was obtained by firstly converting the measured EL spectra to power spectral densities $\rho(\lambda)$(W/nm) at different current densities using a calibrated lamp standard (Labsphere AUX-100) and then calculated by the following formula:

$$EQE = \frac{q}{hc}\int \rho(\lambda)\lambda d\lambda$$

where q is the electric charge, h is Planck constant, c is light speed in air and $\lambda$ is the wavelength.

The measured EQE shows that in the current density range from 1mA/cm² to 100mA/cm², the HDNM-OELD exhibits a maximum EQE of 48% (at current density <2mA/cm²), which is 1.85-fold higher than that of the ITO-OELD (a maximum EQE of 26%). And the EQE of 48%, to our best knowledge, is the highest achieved for green emitting OELDs using CBP: Ir(ppy)$_3$ as the host: guest material in the emission layer. Considering $EQE = IQE \times \eta_{ex.}$ where $\eta_{ex.}$ is the light extraction efficiency, and we assume the IQE of the light emitting materials is not affected by the different OELD structures, hence the light extraction efficiency enhancement should be the same as the EQE enhancement. Therefore, based on the measured EQE, the light extraction enhancement for this green-emitting HDNM-OLED structure is 1.85-fold.

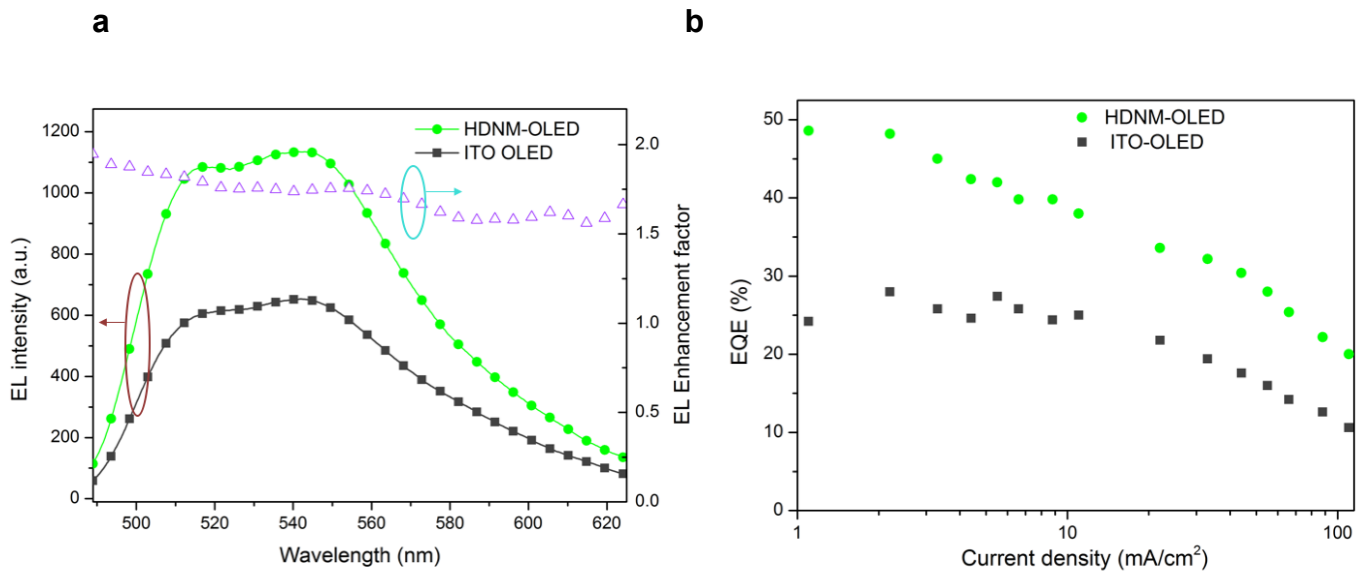

**Figure 6. Measured electroluminescence (EL) and EQE.** (a) Total front-surface EL & EL enhancement spectrum(b) EQE vs. current density. Compared with ITO-OLEDs, HDNM-OELDs show 1.71-fold average EL enhancement factor and 1.85-fold maximum EQE enhancement factor.

**Current-density-voltage characteristic, maintain good charge transport property and power efficiency enhancement.** The current density vs. operating voltage (J-V) characteristics were measured for both the HDNM-OLED and the ITO-OLED using a source meter (HP 4145B). As shown in the Fig. 7a, HDNM-OLED exhibited almost the same J-V behavior as ITO-OLED, indicating the HDNM-OLED structure can maintain the carrier transport property as good as the control device fabricated on planar ITO electrode. Using the measured J-V characteristics and power spectral densities measured at different injection current densities, we obtained the power efficiency as a function of injection current density of HDNM-OLEDs and ITO-OLEDs (Fig.7b) by integrating the power spectral densities with human eye's luminosity function and dividing by the input electrical power. Compared with the ITO-OLED, the maximum power efficiency of the HDNM-OLED was achieved at 1mA/cm$^2$ and exhibited 1.86-fold enhancement increasing from 42lm/W to 79lm/W. Here we find that the maximum power efficiency enhancement factor (1.86-

fold) is almost the same as the maximum EQE enhancement factor (1.85-fold) which is reasonable. The power efficiency is the ratio of outcoupled luminous power to the input electrical power and hence the power efficiency is a combined effect of both light extraction performance and J-V characteristic. For HDNM-OELD, because the charge transport property was maintained as same as the ITO-OLED, only the light extraction enhancement contributed to the overall power efficiency enhancement. In our previous study of the DNM-OLED structure, we found that making the OLED organic layers corrugated by directly depositing the organic layers and electrodes on the nanostructure patterned substrate can even improve the charge transport property by reducing the driving voltage at a given current density. But that is at the cost of limiting the ability of achieving high light extraction efficiency since thin organic layers cannot sustain large surface roughness of the substrate which is required for achieving high scattering efficiency and hence high light extraction efficiency. For DNM-OLED in our previous study, the maximum light extraction efficiency enhancement factor is only 1.33-fold but the maximum power efficiency enhancement factor is 1.8-fold, from which we can see the improved charge transport played a big role in improving the overall power efficiency. The new HDNM-OLED we designed in this paper can achieve significantly high light extraction enhancement without degrading the charge transport property of the OLED which solved the big problem of the previous DNM-OLED structure. But if we can do better than maintaining the charge transport property for our HDNM-OLED by adopting the corrugation organic layer structure in DNM-OLED to improve the charge transport property, we should be able to further increase the current 1.86-fold power efficient enhancement factor. A possible way should be using a weaker top planarization layer to make the top surface of the HDNM-substrate have appropriate surface corrugation roughness that the organic layers can handle. But if it can be easily controllable is still a question.

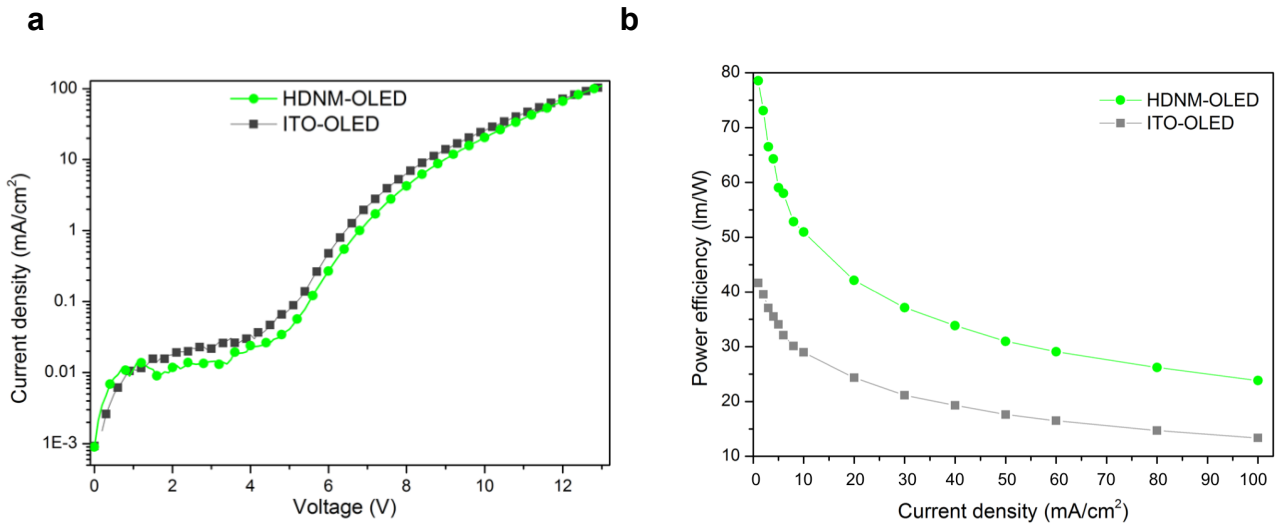

**Figure 7. Measured J-V characteristic and power efficiency of HDNM-OLEDs and ITO-OLEDs.** (a) current density vs. driving voltage (b) power efficiency vs. current density. HDNM-OLEDs and ITO-OLEDs show almost the same J-V characteristic. Compared to ITO-OLEDs, DNM-OLEDs show 1.86-fold maximum power efficiency enhancement.

**Emission profiles and angular dependence of luminous intensity.** The 2D far field emission patterns of the ITO-OLED and the HDNM-OLED was obtained be taking the optical images of the operating OLEDs in normal direction using CCD camera (Fig. 8a). It clearly shows that the ITO-OLED exhibited an isotropic emission pattern whereas the HDNM-OLED exhibited stronger emission intensities along those two periodic directions of the nanomesh nanostructure because of the constructive interference of Bragg scattering. And along the periodic direction, we measured the luminous intensity vs. the emission angle of both ITO-OLED and HDNM-OLED. The angular dependent luminous intensities were normalized to the intensity measured at normal direction as shown in Fig. 8b. The HDNM-OLED showed the strongest emission enhancement at the emission angle around 60 degree which is very different from the typical lamebrain emission profile of the ITO-OLED. This angular dependent emission enhancement can be explained by the Bragg's law $k_0 \sin\theta = |\mathbf{k}_{trap} \pm \mathbf{\Lambda}|$, where $\theta$ is the emission angle, $\mathbf{k}_{trap}$ is the lateral wave vector of the photons trapped in the OLED, $k_0$ is the amplitude of the wave vector of the free photons in air and $\mathbf{\Lambda}\left(\frac{n_x 2\pi}{a}\mathbf{i} + \frac{n_y 2\pi}{a}\mathbf{j}\right)$, where a is the pitch of nanomesh) is the lattice momentum. At around 60 degree, the wavelength satisfying the Bragg's law just matches the emission peak wavelength of green emitting the HDNM-OLED.

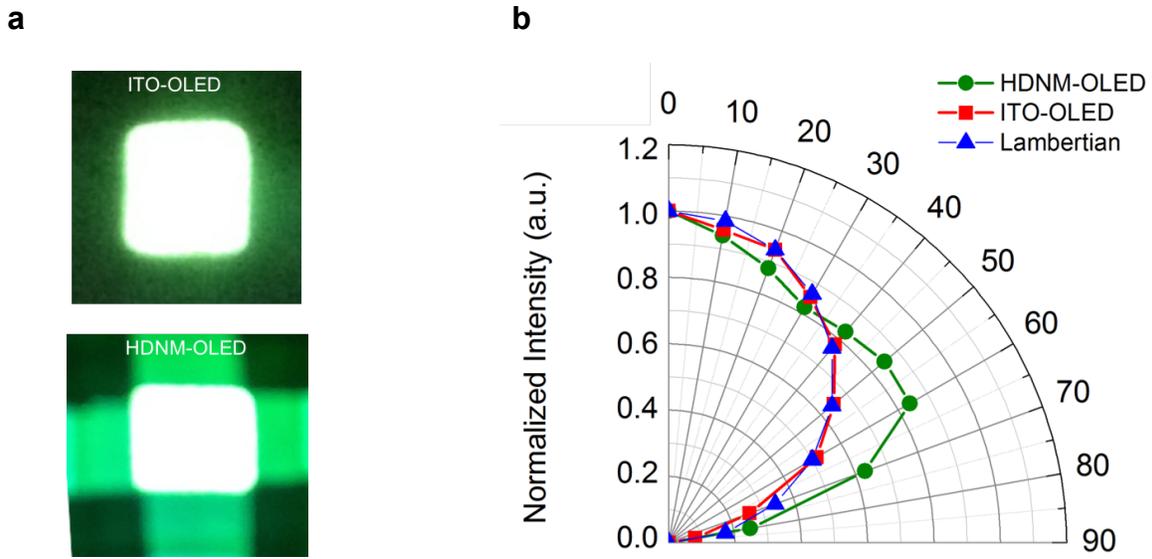

**Figure 8. HDNM-OELD structure modify the emission profile** (a) Optical images of the operating ITO-OELD (upper) and HDNM-OLED (bottom) showing the 2D far-filed emission pattern (b) normalized luminance vs. emission angle of ITO-OLEDs and HDNM-OLEDs (along the periodic direction of nanomesh).

**Groove depth affect light extraction enhancement.** As discussed above, the higher the scattering efficiency of the nanostructure is, the higher light extraction enhancement the OLED can achieve. Because the groove depth significantly affect the scattering ability of the nanomesh, we experimentally investigated how the groove depth of the nanomesh affect the light extraction efficiency of the HDNM-OLED. A series of same HDNM-OELD devices except different nanomesh groove depths (60nm, 125nm, 300nm and 400nm) were fabricated, and the maximum EQEs of those devices were measured and plotted in Fig. 9 as relative light extraction efficiencies with respect to the maximum EQE of the ITO-OLED. It is shown in Fig. 9 that the light extraction efficiency of the HDNM-OLEDs monotonically increases to around 1.7-fold with the increasing groove depth until reaching saturation at the groove depth of around 300nm, which is consistent with the Mie scattering theory as shown in Fig. 2: the scattering cross section of a scatterer gradually increases with the increasing size and then saturates. And this consistence further confirmed that the HDNM-OLED structures enhance the light extraction without electrical properties degradation since the light extraction performance of the HDNM-OLED only depends on the scattering performance of the nanomesh patterned HDNM-substrate. If recalling our previous reported DNM-OLED structure, the light extraction enhancement factor of DNM-OLEDs began to decrease from the highest value of 1.3-fold as soon as the groove depth just increased above 40nm. Comparing the groove-depth dependence of the light extraction efficiency between HDNM-OLED and DNM-OLED, we can see that to achieve a high light extraction enhancement factor, maintaining a good charge transport property is very important to the method using nanostructured substrate for light extraction from OLEDs.

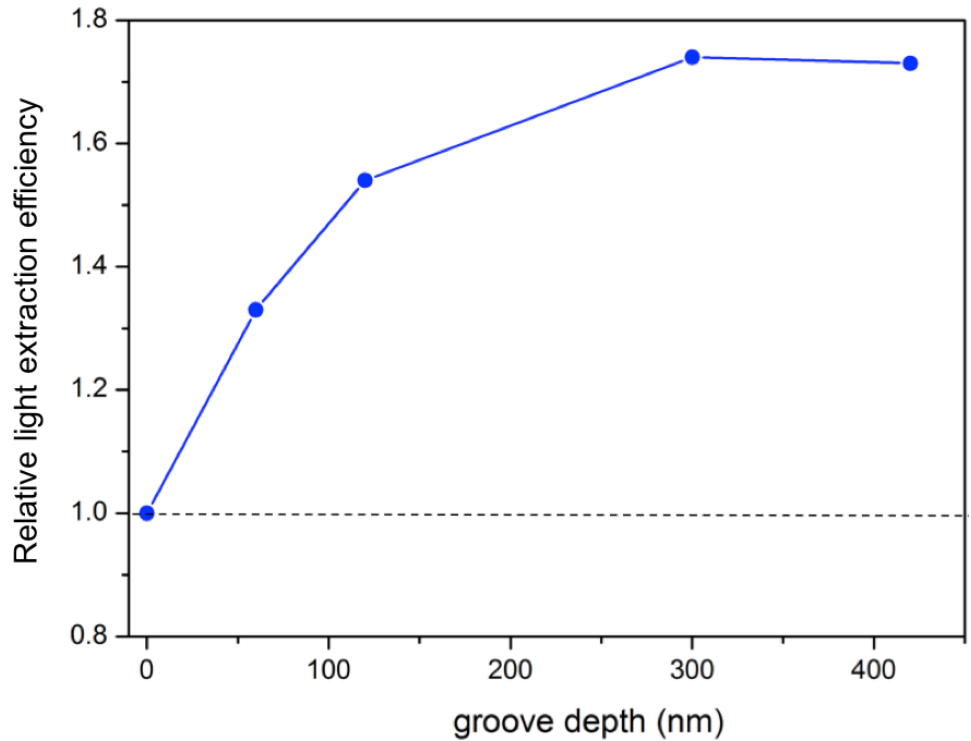

**Figure 9. Relative light extraction efficiency of HDNM-OLEDs with respect to the ITO-OLEDs as a function of the groove depth of the nanomesh.**

**Red and blue light emitting HDNM-OELD.** As we discussed in our previous work of DNM-OLED, the pitch size of the nanomesh is properly chosen to be 400nm so that the photons of a broad range of wavelength from 320nm to 800nm covering the whole visible wavelength can satisfy the Bragg law and get extracted from the OLED. Hence, to experimentally demonstrate the HDNM-OLED can achieve broadband light extraction enhancement in the entire visible light range, we fabricated red and blue-emitting HDNM-OELDs and the corresponding ITO-OELDs and measured the EL spectra and EQE as a function of the injection current (Fig. 10). The red and blue-emitting HDNM-OLEDs have the same device structure as the green-emitting DNM-OLEDs discussed above except the light emitting layers. The light emitting layers of the red-emitting OLED are NPB (40nm)/CBP: Ir(piq)$_2$acac (6wt%)(30nm)/TPBI (50nm). Bis(1-phenylisoquinoline)(acetylacetonate)iridium(III) [Ir(piq)2acac] is the high efficient phosphorescent dopants that emit red light (emission peak: 628nm). The light emitting layers for blue-emitting OLED are NPB (40nm)/mCP (20nm)/SPPO1:FIrpic (10wt%) (30nm)/SPPO1 (40nm), where 1,3-Di(9H-carbazol-9-yl)benzene (mCP) is the exciton blocking material and 9,9-Spirobifluoren-2-yl-diphenylphosphine oxide (SPPO1) is both the host material for high efficient blue emitting phosphorescent dopants Bis[2-(4,6-difluorophenyl)pyridinato-C2,N] - (picolinato)iridium(III) [FIrpic] (emission peak: 475nm~525nm ) and electron transport material.

The measured EL spectra show that, compared to ITO-OLED, the red-light emitting HDNM-OELD exhibited 1.49-fold average EL enhancement (Fig. 10a) and the blue-light emitting HDNM-OLED exhibited 1.72-fold average EL enhancement (Fig. 10c). The measured EQE vs. injection current density shows that, compared with the ITO-OLED, the maximum EQE of the HDNM-OELD (at 1mA/cm$^2$) exhibited 1.58-fold enhancement increasing from 13% to 21% for red light (Fig. 10b) and 1.86-fold enhancement increasing from 22% to 41% for blue light (Fig. 10d). Combined with the result of EQE enhancement by HDNM-OLED structure for green-emitting OLEDs as shown above, it has been experimentally demonstrated that the HDNM-OLED structure is well designed for broadband light extraction enhancement over the entire visible wavelength range.

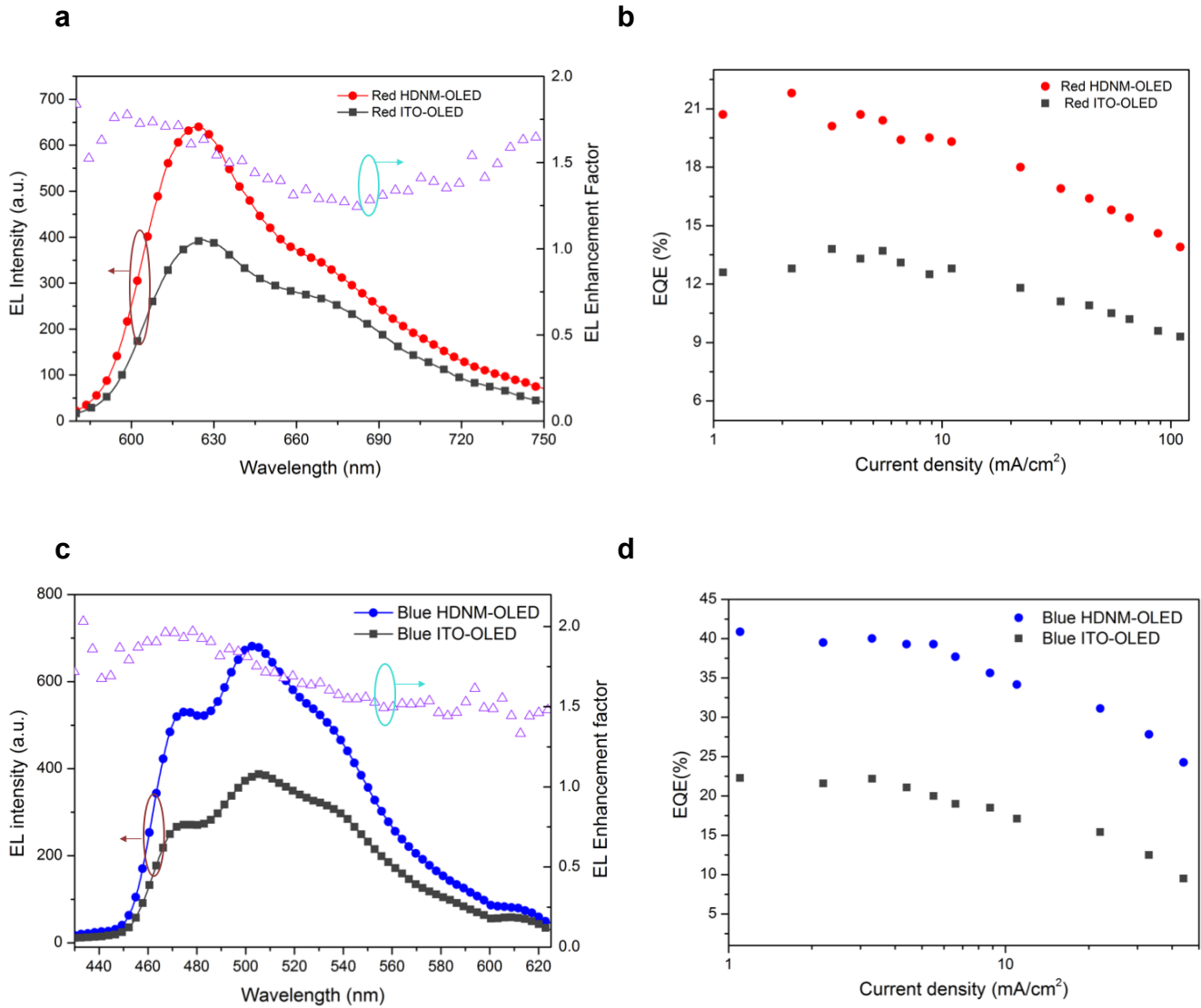

**Figure 10. Measured EL spectra and EQE of red and blue HDNM-OLEDs.** EL spectra of red (a) and blue (c) HDNM-OLEDs. EQE vs. current density of red (c) and blue (d) HDNM-OLEDs. Compared with the ITO-OLLEDs, the red emitting HDNM-OLED exhibits 1.49-fold average EL enhancement and 1.58-fold maximum EQE enhancement increasing from 13% to 21% and the blue emitting HDNM-OLED exhibits 1.72-fold average EL enhancement and 1.86-fold maximum EQE enhancement increasing from 22% to 41% for blue light.

One interesting observation here is that although HDNM-OLED structure can enhance the light extraction efficiency of all red, green and blue-emitting OLED, the enhancement factor of red-emitting HDNM-OLED is much lower that of the green or blue-emitting HDNM-OLED (1.58–fold vs. 1.85-fold). This phenomenon can be explained by considering Bragg scattering law and the angular dependent Mie scattering intensity. Compared to the 400nm pitch size of the nanomesh, the emission wavelength of the red-emitting OLED (emission peak: 628nm) has a larger difference than that of the green and blue-emitting OLEDs. Hence, it requires large angle scattering for the trapped red-emitting photons to satisfy the Bragg law and get extracted. And according the Mie scattering theory, the scattering intensity of a scatterer is very angular dependent as illustrated in the Fig. 11: the scattering intensity of a small angle scattering is generally much larger that of a large angle scattering[21,32-34]. Taking these into consideration, it is reasonable that the measured light extraction enhancement factors of green and blue-emitting HDNM-OLEDs are much higher that of the red-emitting HDNM-OELD.

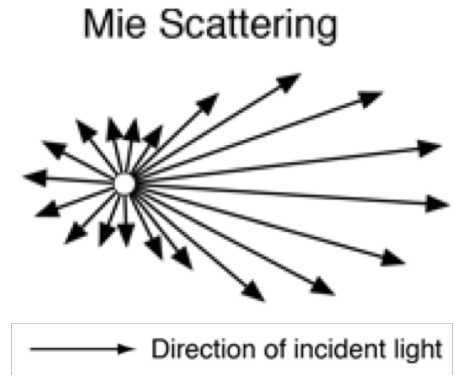

**Figure 11. Illustration of the angular dependence of Mie scattering intensity.** Small angle scattering has larger scattering intensity than that of a large angle scattering.

**Conclusion**

In summary, we developed a novel OLED structure, HDNM-OLED, containing a newly designed HDNM-substrate patterned with high-index deep-groove dielectric-nanomesh. Experimentally, the HDNM-OLEDs showed significantly high light extraction enhancement without introducing electrical property degradation. On the one hand, the HDNM-OLEDs exhibited high broadband light extraction efficiency enhancement. Compared with the ITO-OLEDs, the HDNM-OELDs showed maximum EQE increased from 26% to 48% for green light (1. 85-fold enhancement factor), from 22% to 41% for blue light (1.86-fold enhancement factor) and from 13% to 21% for red light (1.58-fold enhancement factor). The light extraction enhancement is because of the Bragg scattering of the trapped photons by the nanomesh structures. And the high refractive-index contrast within the HDNM-substrate ensures a high Mie scattering efficiency which is essential for achieving high light extraction enhancement factor. On the other hand, the HDNM-substrate does not modify the morphology of the OLED organic active layers deposited on top because of the top planarization layer so that the charge transport property of HDNM-OLEDs is as good as that of a conventional ITO-OLED. In sum, this novel HDNM-OLED structure solved the conventional key conflict between maintaining good charge transport property and achieving high light extraction efficiency when using micro/nanostructure patterned substrates to extract light from OLEDs. Moreover, another advantage of the HDNM-OLEDs is the easy and scalable fabrication by nanoimprint lithography. With improvements in light emitting materials and optimization of structure design, the above HDNM-OLEDs' performances can be further improved. The designs, fabrications, and findings are applicable to the LEDs in other materials (organic or inorganic) and on other thin substrates (plastics or glasses), thus opening up new avenues in developing high efficient OLED lighting and display.